\documentclass[12pt,amsfonts]{article}   
\usepackage{amsfonts}
\def\t{}                 
\def\one{1\hskip-.37em 1}                 

\def\half{{\textstyle{\frac{1}{2}}}}

\def\op{\overrightarrow{p}}
\def\oq{\overrightarrow{q}}
\def\oP{\overrightarrow{P}}
\def\oQ{\overrightarrow{Q}}
\def\orr{\overrightarrow{r}}
\def\os{\overrightarrow{s}}
\def\oR{\overrightarrow{R}}
\def\oS{\overrightarrow{S}}
\def\b{\beta}

\def\H{{\cal H}}

\def\tilde{\widetilde}
\def\H{{\cal H}}

\def\l{\lambda}

\def\ra{\rightarrow}

\def\tint{{\textstyle\int}}

\def\s{\hskip.08em}
\def\d{\partial}

\def\b{\begin{eqnarray*}}  
\def\e{\end{eqnarray*}}    
\def\bn{\begin{eqnarray}}  
\def\en{\end{eqnarray}}   
\def\<{\langle}
\def\>{\rangle}

\def\no{\nonumber}

\def\{{\lbrace}
\def\}{\rbrace}

\title{Revisiting  Canonical Quantization}    
\author{John R. Klauder\footnote{Email: klauder@phys.ufl.edu}\\
Department of Physics and\\Department of Mathematics\\
University of Florida\\
Gainesville, FL 32611-8440}
\date{ }

\begin{document}
\maketitle
\begin{abstract}
Conventional canonical quantization procedures directly link various $c$-number and $q$-number
quantities. Here, we advocate a different association of classical and quantum quantities that renders
classical theory a natural subset of quantum theory with $\hbar>0$, in conformity with the real world  wherein  nature has chosen $\hbar>0$ rather than $\hbar=0$. While keeping the good results of conventional procedures, some examples  are presented for which the new procedures
offer better results than conventional ones.
\end{abstract}
\section{Introduction}The most common approach to a quantum theory is through Schr\"odinger's equation
    \bn i\hbar\s\s\partial\s\psi(x,t)/\partial t={\cal H}(-i\hbar\s\s\partial/\partial x,x)\,\psi(x,t) \;, \label{es}\en
    illustrated for a single degree of freedom. Here the function ${\cal H}(p,q)$ generally differs from   the classical ($c$) Hamiltonian $H_c(p,q)$ by terms of order $\hbar$, the coordinates $p\ra -i\hbar\partial/\partial x$ and $q\ra x$, and
$\int_{-\infty}^\infty|\psi(x,t)|^2\,dx=1$. A similar prescription applies to classical systems
with $N$ degrees of freedom, $N\le\infty$. Although this scheme is widely successful, there are certain
questionable aspects. Generally, the given procedure works well only for certain canonical coordinate systems,
namely, for ``Cartesian coordinates'' \cite{dir}, despite the fact that the classical phase space
has a symplectic structure (e.g., $dp\; dq$, interpreted  as an element of surface area) but {\it no} metric structure
(e.g., $\omega^{-1} dp^2 +\omega\s\s dq^2$) \cite{ham} with which to identify Cartesian coordinates. Moreover, for certain classical systems, and even when
using the correct coordinates to provide a canonical quantization, a subsequent classical limit in which $\hbar\ra0$ leads to a manifestly {\it different}
classical system from the original one, thus violating the eminently natural rule that de-quantization should
lead back to the original classical system. Finally, the classical framework for which $\hbar=0$ is
fundamentally different from the quantum framework for which $\hbar>0$, and it is the latter realm that
characterizes the real world.

There are several arguments that support a new approach. First, by analogy, note that the real world is relativistic in character, but a nonrelativistic approximation to classical mechanics
can be made within relativistic classical mechanics {\it without changing the formulation} and keeping the speed of light $c$ fixed and finite. Likewise, since the real world is also governed by quantum mechanics, it is necessary that classical mechanics somehow be contained within quantum mechanics in such a way that {\it it involves the same formulation} and keeps the reduced Planck's constant $\hbar=h/2\pi$ fixed and nonzero. Second, the current prescription on how a classical system should be quantized leads, for some
problems---e.g., self-interacting scalar fields
$\varphi^4_n$, for spacetime dimensions $n\ge5$ (and possibly $n=4$ as well)---to unnatural behavior in that the classical limit of a conventional quantization does {\it not} reduce to the original classical model when $\hbar\ra0$! As we shall illustrate, this serious discrepancy can be overcome with procedures analogous to those discussed in the
present article. However, we do not need to invoke complicated examples to learn some of the advantages of a new
way to link classical and quantum systems together.

\section{Enhanced Canonical Quantization}
As discussed above, we shall propose a different manner of quantization---called {\it Enhanced
Quantization}---that {\it keeps all the good results of conventional canonical quantization, but offers
better solutions when needed}. We start with the quantum action functional $A_Q$ given, for a single degree of freedom
and normalized wave functions, by
       \bn A_Q=\tint^T_0\tint^\infty_{-\infty}\s\psi^*(x,t)\s[\s i\hbar\s\partial/\partial t- {\cal H}(-i\hbar\partial/\partial
x,x)\s]\s\psi(x,t)\,dx\,dt \label{aq} \en
       from which Schr\"odinger's equation (\ref{es}) may be derived by a general stationary variation,
$\delta A_Q=0$, provided that
       $\delta\psi(x,0)=0=\delta\psi(x,T)$. Such variations correspond to variations that can actually be realized in practice, but,
       in some situations, not all variations are possible. As an example,
       consider a {\it microscopic system}: then while {\it microscopic observations} can make sufficiently general  variations to
       deduce Schr\"odinger's equation, {\it macroscopic observations} are confined to a much smaller
subset of possible variations. For example, we include only those that can be realized {\it without disturbing the observed system}, such as changes made in accordance with Galilean invariance, namely a change in position  by $q$ and a change in momentum by $p$ (as realized by a change in velocity); note that no disturbance of the observed system need occur since we can instead translate the reference system. Choosing a foundation on which to build, we select a basic
normalized function---call it $\eta(x)$---which, when transported by $p$ and $q$ as noted above, gives rise to a family of functions $\eta_{p,q}(x)\equiv e^{ip(x-q)/\hbar}\s\eta(x-q)$, where $-\infty<p,q<\infty$ [note: $\eta_{\s0,0}(x)\equiv\eta(x)\s$]. Within the context of quantum mechanics, these functions are also well known as {\it canonical coherent
states} \cite{klsk}, and the basic function is generally referred to as the fiducial vector, all expressed here in the Schr\"odinger representation. While not required, it is useful to impose $\tint_{-\infty}^\infty x|\eta(x)|^2dx =0$ as well as
$\tint_{-\infty}^\infty \eta(x)^*\s\eta'(x)\s dx=0$, called ``physical centering'', which then leads to
  \bn \tint_{-\infty}^\infty x\s|\eta_{p,q}(x)|^2\,dx=q\,,\hskip2em -i\hbar\tint_{-\infty}^\infty
\eta_{p,q}(x)^*\s\eta'_{p,q}(x)\,dx=p \;,\en
two relations that fix the physical meaning of $p$ and $q$, independently of the basic function $\eta(x)$.
Finally, being unable as macroscopic observers to vary the functions $\psi(x,t)$ in (\ref{aq}) arbitrarily, we restrict (R) the set of allowed variational states so that
$\psi(x,t)\ra \eta_{p(t),q(t)}(x)$, which leads to
         \bn A_{Q(R)}\hskip-1.3em&&=\tint^T_0\tint^\infty_{-\infty}\s\eta_{p(t),q(t)}(x)^*\s[i
\hbar\s\partial/\partial t- {\cal H}(-i\hbar\partial/\partial x,x)\s]\s\eta_{p(t),q(t)}(x)\,dx\,dt\no\\
           &&=\tint^T_0\s[p(t)\s {\dot q}(t)-H(p(t),q(t))\s]\,dt\;,  \label{e3} \en
           where
\bn H(p,q)\equiv \tint_{-\infty}^\infty \s\eta(x)^* {\cal H}(p-i\hbar\partial/\partial x,q+x)\s\s
\eta(x)\,dx\;. \label{e4}\en
Because $\hbar>0$ still, we call (\ref{e3}) and (\ref{e4}) the {\it enhanced classical action functional}.
Assuming, for simplicity,  that the Hamiltonian operator is a polynomial, it readily follows that
   \bn H(p,q)={\cal H}(p,q)+{\cal O}(\hbar;p,q) \label{ohbar}\en
for many choices of the basic function $\eta(x)$. In such cases, the strictly classical action functional arises if the limit $\hbar\ra0$ is applied to (\ref{e3}) and (\ref{e4}).

       The introduction of classical canonical coordinate transformations---for example,
$(p,q)\ra ({\tilde p},{\tilde q})$ such that
               $p\s\s dq={\tilde p}\s\s d{\tilde q}+d{\tilde G}({\tilde p},{\tilde q})$,
and for which we choose  ${\tilde\eta}_{{\tilde p},{\tilde
          q}}(x)\equiv\eta_{p({\tilde p},{\tilde q}),q({\tilde p},{\tilde q})}
(x)=\eta_{p,q}(x)$---leads to an enhanced classical action functional properly expressed
in the new coordinates, but with {\it no} change of the quantum formalism---nor of the
physics---whatsoever.

         Stationary variation of the enhanced classical action functional leads to Hamilton's equations
of motion based on the enhanced classical Hamiltonian $H(p,q)$. In this
sense we have  shown that a suitably restricted domain of the {\it quantum
action functional}, consisting of a two-dimensional, continuously connected sheet of normalized
functions,  leads to an enhanced {\it canonical classical action functional}, with the benefit that $\hbar>0$
still, and, in this way, we have achieved the goal of {\it embedding classical mechanics within quantum mechanics!}
Moreover, the enhanced classical equations of motion may have ${\cal O}(\hbar)$ correction terms---the form of which
could be dictated by (\ref{e4})---that
may be of interest
in modifying the strictly classical solutions in interesting ways. One example of this behavior is offered below

In a crude sense, and for a reasonable range of $(p,q)$, elements of the set of states $\{\eta_{p,q}(x)\}$ act like the illumination from a flashlight
used by a  burglar
in peering through the window of a deserted house on a pitch black night; indeed, the role of physically
centering the
function $\eta(x)$ is like ensuring the flashlight is aimed through the window and not at the brick wall.
Changing the basic function $\eta(x)$
is like changing the orientation or the cone of illumination of the flashlight. Quantum mechanically, whatever the choice of
$\eta(x)$, the set of  functions
$\{\eta_{p,q}(x)\}$, for all $(p,q)$, span the space of all square integrable functions, $L^2({\mathbb R})$. In further analysis of
the enhanced classical theory, however, some choices of $\eta(x)$ may  be better than others.

A common choice for $\eta(x)$---and one which acts like a bright, narrow-beam flashlight---is given, with  $\omega>0$,  by
  \bn \eta(x)=(\omega/\pi\hbar)^{1/4}\s e^{-\omega x^2/2\hbar} \label{f9}\en
for which  $H(p,q$) satisfies (\ref{ohbar}), i.e., ${\cal H}(p,q)=H(p,q)$ up to terms of order $\hbar$. This last property is
{\it exactly} what is meant by having ``Cartesian coordinates'', although such coordinates can not originate from the classical phase space. However, Cartesian
coordinates do have a natural origin from an enhanced quantization viewpoint.
The set of allowed variational states $\{\eta_{p,q}(x)\}$---a set of canonical coherent states as noted
earlier---forms a two-dimensional, continuously connected sheet of normalized functions within the set of  normalized square integrable
functions, and a natural metric can be given for such functions. Since the overall phase of any wave function
carries no physics, the (suitably scaled)  square of the distance  between two
coherent-state rays  is given by
     \bn d^2_R(p',q';p,q)\equiv (2\hbar)\s\min_\alpha\tint_{-\infty}^\infty|\eta_{p',q'}(x)
-e^{i\alpha}\s\eta_{p,q}(x)|^2\,dx\;,\en
     which for two infinitesimally close coherent-state rays becomes (this is also the Fubini-Study metric \cite{fs})
     \bn d\sigma^2(p,q)\equiv(2\hbar)\s[\s\tint |\,d\s\eta_{p,q}(x)|^2\,dx-|\tint\s\eta_{p,q}(x)^*
\,d\s\eta_{p,q}(x)\,dx\,|^2\s]\;, \label{u7}\en
     and for $\eta(x)=(\omega/\pi\hbar)^{1/4}\s\exp(-\omega x^2/2\hbar)$, (\ref{u7}) reduces to
        \bn d\sigma^2(p,q)=\omega^{-1}\s dp^2+\omega\s dq^2\;,\en
        which ensures that $p$ and $q$ are indeed Cartesian coordinates. Although this metric originates in the quantum theory with the canonical coherent states, it may also be assigned to the classical phase space as well.

{\bf Note well:} At this point {\it we have recreated conventional canonical quantization} in that we have identified canonical
variables $p$ and $q$ that behave as Cartesian coordinates, and for which the quantum Hamiltonian is
effectively the same as the conventionally chosen one---particularly for classical Hamiltonians of the form $p^2/(2m)+V(q)$---especially if $\omega$ in (\ref{f9}) is chosen very large. {\bf Thus, enhanced quantization can reproduce
conventional canonical quantization---{\it but it has other positive features as well!}}

\section{Enhanced Affine Quantization}
Classical canonical variables $p$ and $q$ fulfill the Poisson bracket $\{q,p\}=1$, which translates to the
Heisenberg commutation rule $[x,-i\hbar\s(\partial/\partial x)]=i\hbar\s$; these operators generate the two transformations that characterize the canonical coherent states.
Multiplying the Poisson bracket by $q$ leads to $\{q,p\s q\}=q$, which corresponds to
$-(i\hbar/2)\s[x,x\s(\partial/\partial x)+(\partial/\partial x) x]=i\hbar\s x$ after both sides of the commutator are multiplied by $x$. This expression, {\it derived from the Heisenberg commutation relation}, is known as an affine commutation relation between affine variables. While the operator $-i\hbar\s(\partial/\partial x)$ acts
to generate {\it translations} of $x$, the operator $-(i\hbar/2)[x\s(\partial/\partial x)+(\partial/\partial x)\s x]$ acts to generate {\it dilations} of $x$. If one deals with a classical variable $q>0$ and its quantum analog $x>0$ (both chosen dimensionless for convenience), then the canonical coherent states are unsuitable and we need a different set of coherent states. We choose a new basic function $\xi(x)\equiv M\s x^{{\tilde\beta}/\hbar-1/2}\,e^{-{\tilde\beta}\s x/\hbar}$, ${\tilde\beta}>0$ and $x>0$, with $M$ a normalization factor, for which it also follows that $\tint_0^\infty x|\xi(x)|^2\,dx=1$ and $\tint_0^\infty \xi(x)^*[x(\partial/\partial x)+(\partial/\partial x)x]\xi(x)\,dx=0$. We also introduce suitable {\it affine coherent states} as $\xi_{p,q}(x)\equiv q^{-1/2}e^{ip(x-q)/\hbar}\,\xi(x/q)$, where $q>0$ and $-\infty<p<\infty$
[note: $\xi_{\s0,1}(x)\equiv\xi(x)\s$, and the phase factor $e^{-ip\s q/\hbar}$ is normally omitted in defining affine coherent states].
The physical meaning of the variables $p$ and $q$ follows from the relations
   \bn  q=&&\hskip-1.3em\tint_0^\infty \xi_{p,q}(x)^*\s x\s\xi_{p,q}(x)\,dx\;,\no\\
      p\s\s q=&&\hskip-1.3em-\half i\hbar\tint_0^\infty \xi_{p,q}^*(x)\s [x(\d/\d x)+(\d/\d x)x)\s]\s\xi_{p,q}(x)\,dx\;. \en
The affine coherent states involve translation in Fourier space by $p$ and dilation---i.e., (de)magnification,  partially realized by a magnifying glass---in configuration space by $q$, and they describe a new, two-dimensional, continuously connected sheet of normalized functions.

The quantum action functional on the half space $x>0$ is given by
\bn  A'_Q=\tint_0^T\tint_0^\infty\,\phi(x,t)^*\s[\s i\hbar\partial/\partial t-{\cal H}(-i\hbar\s\partial/\partial x,x)\s]\s\phi(x,t)\,dx\, dt \;,\en
and a suitable stationary variation leads to Schr\"odinger's equation. Restricted (R) to the affine coherent states, we find that
  \bn  A'_{Q(R)}\hskip-1.3em&&=\tint_0^T\tint_0^\infty\,\xi_{p(t),q(t)}(x)^*\s[\s i\hbar\partial/\partial t-{\cal H}(-i\hbar\s\partial/\partial x,x)\s]\,\xi_{p(t),q(t)}(x)\,dx \,dt \no\\
 &&=\tint_0^T[\s p(t)\s{\dot q}(t)-H(p(t),q(t))\s]\,dt\;,  \label{f3} \en
where
   \bn H(p,q)\equiv \tint_0^\infty\s\xi(x)^*\s{\cal H}(p-iq^{-1}\hbar\partial/\partial x,q\s x)\,\xi(x)\,dx\;.
\label{f4}  \en
Equations (\ref{f3}) and (\ref{f4}) strongly suggest that they correspond to an enhanced {\it canonical  classical action functional}. In other words, enhanced quantization has found a {\it different two-dimensional sheet of normalized functions} that nevertheless exhibits a {\it conventional canonical system for its
enhanced classical behavior!} Invariance under canonical coordinate transformations follows along the same lines as before. For this system, ``Cartesian coordinates'' are {\it not} appropriate; instead the geometry of the affine coherent-state rays leads to a Fubini-Study metric given by
  \bn d\sigma^2(p,q)\hskip-1.3em&&\equiv(2\hbar)\s[\s \tint|\,d\s\xi_{p,q}(x)|^2\,dx-|\s\tint\s\xi_{p,q}(x)^*\s d\s\xi_{p,q}(x)\,dx\s|^2\s]\no\\
  &&={\tilde\beta}^{-1}\s q^2\,dp^2+{\tilde\beta}\s q^{-2}\,dq^2\;, \en
which is a space of constant negative curvature:$-2/{\tilde\beta}$ (a Poincar\'e half plane \cite{PO}). As with the canonical case,
this new geometry can be added to the classical phase space if so desired.

While we have focussed on the case where $x>0$, it is also possible to consider $x<0$, and
even a {\it reducible representation} case where $|x|>0$, in which case, enhanced affine quantization can, effectively, replace enhanced canonical quantization.

\section{Examples}
\subsection{Model one}
Consider the classical action functional for a single degree of freedom given by
   \bn A_C=\tint^T_0[\s p\s{\dot q}- q\s p^2\s]\,dt\;, \en
   with the physical requirement that $q>0$. The classical solutions for this example are given by
     \bn p(t)=p_0\s(1+p_0\s t)^{-1}\;,\hskip2em q(t)=q_0\s(1+p_0\s t)^2\;, \en
     where $(p_0,q_0)$ denote initial data at $t=0$. Although $q(t)$ is never negative,
     {\it every solution} with positive energy, $E=q_0\s p_0^2>0$, becomes singular since $q(-p_0^{-1})=0$.

     We like to believe that quantization of singular classical systems may, sometimes, eliminate the
     singular behavior, and let us see if that can occur for the present system. Conventional canonical quantization is ambiguous up to terms of order $\hbar$, and that makes it difficult to decide
     on proper nonclassical corrections when $\hbar>0$. In contrast, enhanced quantization
     always keeps $\hbar>0$ and points to quite specific nonclassical corrections.
     Adopting enhanced affine quantization, the enhanced classical action functional becomes
       \bn A_{Q(R)}\hskip-1.3em&&=\tint^T_0\tint^\infty_0\xi_{p(t),q(t)}(x)^*\s[\s i\hbar\partial/\partial t-(-i\hbar\partial/\partial x)\s x\s(-i\hbar\partial/\partial x)\s]\s\xi_{p(t),q(t)}(x)\,dx\no\\
         &&=\tint_0^T[p(t)\s{\dot q}(t)-q(t)\s p(t)^2-C\s q(t)^{-1}\s]\,dt\;, \en
         where  $ C\equiv \hbar^2\,\tint_0^\infty x\s |\xi'(x)|^2\,dx>0$.
          With our convention that $q$ and $x$ are dimensionless, it follows that the dimensions of $C$ are those of $\hbar^2$.
          While the numerical value of $C$ may depend on $\xi(x)$, the modification of the classical Hamiltonian has just one term proportional to $\hbar^2 \s q^{-1}$, which guarantees that the enhanced classical solutions do indeed eliminate the divergences encountered in the strictly classical solutions.

          Model one, which represents a toy model of gravity, is based on {\cite{kl3}, which also includes additional details. Recently, Fanuel and Zonettti \cite{fz} have employed affine quantization to study more refined cosmological gravity models, and have concluded that conventional classical singularities are removed upon quantization.

   \subsection{Model two}
   This example involves another feature of enhanced quantization besides that of affine quantization. Although involving more degrees of freedom than model one, which makes it more advanced than the previous model, the present example illustrates an important feature of enhanced quantization. In particular, model two involves
   many (possibly, infinitely many) degrees of freedom: $N\le\infty$. When $N=\infty$, it also provides a valid quantization of a system that {\it fails to be properly quantized} by conventional quantization techniques. The model in question has a classical
   Hamiltonian given by
     \bn H(\op,\oq)=\half[\op^2+m_0^2\s\oq^2]+\l_0\s(\oq^2)^2\;, \label{basic} \en
     where $\op=\{p_1,p_2,\ldots,p_N\}\in{\mathbb{R}}^N$, $\op^2\equiv \op\cdot\op=\Sigma_{n=1}^Np_n^2$  (and likewise for $\oq$); when  $N=\infty$, it is necessary that $\op^2+\oq^2<\infty$.

     To understand why this model is important, we start with a few remarks about canonical quantization of this model when $N=\infty$.
     Canonical quantization of this model when $N=\infty$ leads to
     a {\it free quantum theory} with no real quartic interaction, which clearly passes to a {\it free classical
     model} as $\hbar\ra0$, a limit that is  {\it different} from the original nonfree classical model (\ref{basic}). This unnatural behavior happens because, as $N$ becomes large, nascent divergences caused by the interaction term must be tamed by rescaling $\l_0\ra\l_0/N$, eventually nullifying any real influence of the interaction.

     We may arrive at the same conclusion from a different line of reasoning. The difficulties for this model arise when $N=\infty$, but they can be analyzed and dealt with when $1\ll N<\infty$. We assume that the ground state $\<\overrightarrow{x}|0\>=\psi_0(\overrightarrow{x})$ for this problem is unique
     and has the full rotational symmetry of the model. Thus the characteristic function (i.e., the Fourier transform) of the ground-state distribution becomes
      \bn C(\op)\hskip-1.3em&&=\tint e^{\t i\s\op\cdot\overrightarrow{x}/\hbar} \psi_0(\overrightarrow{x})^2\,d^Nx\no\\
         &&=\tint e^{\t i p_r\s r\s\cos(\theta)/\hbar}\,\rho(r)\,r^{N-1}\,\sin(\theta)^{N-2} dr\s d\theta\,d\Omega_{N-2}\no\\
         &&\simeq \tint e^{\t-p_r^2 \s r^2/2\hbar^2}\,w(r)\,dr \en
         where $p_r\equiv\sqrt{\op^2}$, $r\equiv\sqrt{\overrightarrow{x}^2}$, $d\Omega_{N-2}$ refers to the other angular variables, and the final approximation is  based on a steepest descent evaluation of the $\theta$ integral for large $N$. Thus, in the limit $N\ra\infty$
the most general form of the characteristic function is given by
\bn  C(\op)=\tint_0^\infty e^{\t -b\s \op^2/\hbar}\,d\mu(b) \label{jj} \en
for some normalized, positive measure $\mu$.    
While this analysis deals with preserving full symmetry, the general result in (\ref{jj}) does not respect uniqueness of the desired ground state. To obtain uniqueness of the ground state,
the measure $\mu(b)$ must fulfill $d\mu(b)= \delta(b-1/4m')\s\s db$, which leads to $C(\op)=\exp[-\op^2/4m'\hbar]$, and corresponds to a free system for some positive mass $m'$.

The new approach to this model ensures that the formulation for $N<\infty$ admits a nontrivial limit
as $N\ra\infty$,  a limit that also defines the quantization of the model when $N=\infty$. Consequently, the following analysis primarily assumes that $N<\infty$.
In our discussions below, we mostly use an abstract quantum notation for brevity.
We define coherent states as
  \bn |\op,\oq\>\equiv \exp[-i\oq\cdot\oP/\hbar]\,\exp[i\op\cdot\oQ/\hbar]\,|0\>\;,\en
 where $(m\oQ+i\oP)\s|0\>=0$, and it follows that
  \bn &&\<\op',\oq'|\op,\oq\>\no\\
   &&\hskip-2em =\exp\{i(p'+p)\cdot(q'-q)/2\hbar-[(1/m)(p'-p)^2+m(q'-q)^2 ]/4\hbar\}\;,\en
   an expression ensuring irreducible canonical operators  $\oQ$ and $\oP$, for which $[Q_m,P_n]=i\hbar\s\delta_{m,n}\one$, $1\le m,n\le N$.
   However, for this model, irreducible operators yield a nontrivial result only for $N<\infty$, and so we will need to consider suitable {\it reducible} position and momentum operator representations, for which, effectively, $|0\>$ is replaced by $|0; (K)\>$, such that
   \bn &&\<\op',\oq'; (K)|\op,\oq; (K)\>\no\\
   &&\hskip-2em =\exp\{i(p'+p)\cdot(q'-q)/2\hbar-[(K/m)(p'-p)^2+m(q'-q)^2 ]/4\hbar\}\;, \label{c}\en
   where $K>1$.

            A detailed derivation of a proper quantization of this model is  presented elsewhere; see  \cite{rs1,rs2}. Here, our main goal is to sketch the solution and show that it conforms to enhanced quantization and  not to conventional quantization. To construct suitable reducible operator representations, we
            introduce another, {\it independent}, set of classical and quantum variables, $\orr,\os$ and  $\oR,\oS$, with
            $[S_m,R_n]=i\hbar\delta_{m,n}\one$, $1\le m,n\le N<\infty$. We choose a basic vector $|0; \zeta\>$ given in the Schr\"odinger representation  by
            \bn  \<\overrightarrow{x},\overrightarrow{y}|0; \zeta\>\equiv\psi_{0; \zeta}(\overrightarrow{x},\overrightarrow{y})\equiv M'\,\exp[-m(\overrightarrow{x}^2
            +2\zeta\overrightarrow{x}\cdot\overrightarrow{y}+ \overrightarrow{y}^2)/2\hbar]\;, \en
            where $0<\zeta<1$; the case $\zeta=0$ leads to irreducible operators. Note that our modification is consistent with the ground state being a Gaussian as required by the discussion following (21).
             There are two ``free-looking'' Hamiltonian operators with this vector as their common, unique ground state, namely
            $\H_p\equiv\half:[\oP^2+m(\oQ+\zeta\s\oS)^2]:$ and $\H_r\equiv \half:[\oR^2+m(\oS+\zeta\s\oQ)^2]:$, where $:(\cdot):$ signifies normal ordering with respect to $|0;\zeta\>$.
            We choose new coherent states based on $\psi_{0; \zeta}$ given by
            \bn |\op,\oq; \zeta\>\equiv\exp[-i\oq\cdot\oP/\hbar]\,\exp[\s i\op\cdot\oQ/\hbar]\,|0;\zeta\>\;,\en
            and we consider the operator
           \bn \H_1\equiv \H_p+\H_r+4\nu\s :\H_r^2:\;.\en
            It follows directly that
            \bn\!\!\<\op',\oq';\zeta|\s \H_1|\op,\oq;\zeta\>\hskip-1.3em&&=\{\half[\s (m\oq'-i\op)\cdot(m\oq+i\op)+
               m^2\s\zeta^2\s(\oq'\cdot\oq)\s]\no\\
               &&\hskip-2em+\nu\s\zeta^4\s m^4\,(\oq'\cdot\oq)^2\,\}\,
                \<\op',\oq';\zeta|\op,\oq;\zeta\>\;,\en
               where
               \bn \<\op',\oq';\zeta|\op,\oq;\zeta\>\hskip-1.3em&&\equiv\exp\{\s i(\op'+\op)\cdot(\oq'-\oq)/2\hbar\no\\
               &&\hskip3em-[(\op'-\op)^2/4m(1-\zeta^2)+m\s (\oq'-\oq)^2/4\hbar]\}\no\\
               &&=\exp[i(\op'+\op)\cdot(\oq'-\oq)/2\hbar]\,  \label{eew}\no\\
               &&\hskip-7em\times\s M''(\zeta)\int \, \exp\{i(\op'-\op)\cdot\overrightarrow{y}/\hbar-[(\op'-\op)^2/4m\hbar\no\\
               &&+m\s\hbar\s (\oq'-\oq)^2/4\hbar+m(\zeta^{-2}-1)\overrightarrow{y}^2/\hbar]\}\,d^Ny\;.\en
               In comparison with (\ref{c}), it follows that $K=(1-\zeta^2)^{-1}>1$, and the integral representation in the last line of (\ref{eew}) illustrates a superposition over irreducible representations leading to the desired reducible representation.

               Finally, we observe that
            \bn \<\op,\oq;\zeta|\s \H_1|\op,\oq;\zeta\>\hskip-1.3em&&=\half[\op^2+(1+\zeta^2)\s m^2\s\oq^2]
               +\nu\s \zeta^4\s m^4(\oq^2)^2 \no\\
                 &&\equiv \half[\op^2+m_0^2\s\oq^2]+\l_0\s(\oq^2)^2\;,\en
               {\it which, according to (\ref{basic}), is exactly the expectation value we sought}.

                Yes, conventional quantization requires that the canonical operators need to be realized by an irreducible representation, but that is no longer the case for enhanced quantization since the classical-quantum connection is fundamentally different in the two cases. For enhanced quantization, the classical-quantum connection is generically given by $H(\op,\oq)\equiv\<\op,\oq|\s\H\s|\op,\oq\>$. For the present model, $|\op,\oq\>=|\op,\oq; \zeta\>$, and the classical-quantum connection specifically means that
                   \bn H(\op,\oq)\hskip-1.3em&&\equiv\<\op,\oq; \zeta|\s\H_1(\oP,\oR;\oQ,\oS)\s|\op,\oq; \zeta\>\no\\
                   &&=\half[\op^2+m_0^2\s\oq^2]+\l_0\s(\oq^2)^2\;,\en
                   with no contradiction whatsoever. Note that this feature of enhanced quantization is distinct from affine quantization, and leads to acceptable results for $N\le\infty$. Significantly, when  $N=\infty$, {there are \it no divergences}, unlike the quantization of most interacting models with an infinite number of degrees of freedom.

                   The present models have been studied previously \cite{rs1,rs2}, and some spectral analysis of the interacting Hamiltonian has also been carried out in \cite{rs3}.

\section{Summary and Additional Examples} Conventional canonical quantization, with its direct association between classical variables and quantum variables is correct most of the time, but there are occasions when it leads to unnatural results. Also classical theory assumes that $\hbar=0$ while quantum theory requires that $\hbar>0$; however, in the real world $\hbar>0$ and the classical theory must adapt to that fact, and this is exactly what enhanced quantization achieves. In addition, enhanced quantization offers untapped riches in the very concept of quantization as illustrated by the topic of Sec.~3 and the examples in Sec.~4. Further discussion of these new procedures as well as some additional examples may be found in \cite{kl3,fz,kl6,bg} for some simple models, including an example with spin variables, and \cite{kl7,kll,k33} for several
field-theoretic models---including $\phi^4_n$, $n\ge4$---that, when conventionally quantized, violate the natural rule that the classical limit should result in the original classical theory, but, instead, are successfully
quantized using enhanced quantization procedures. Advantages are found in the study of matrix models \cite{MM}, and in comparing the quantization of bosons, fermions, and anyons \cite{AK}.
 Also analyzed is the kinematics of quantum gravity; see \cite{KL,kl8} and references therein.

\section*{Acknowledgements} Thanks are expressed to  David Tanner for his interest in this work as well
as several useful comments.


\begin{thebibliography}{99}
       \bibitem{dir} P.A.M.~Dirac, {\it The Principles of Quantum Mechanics}, (Clarendon Press,  Oxford, 1947), page 114.

       \bibitem{ham} See, e.g., http://en.wikipedia.org/wiki/Hamiltonian\_mechanics

       \bibitem{klsk} J.R.~Klauder and B.-S.~Skagerstam, {\it Coherent States: Applications to Physics and Mathematical Physics}, (World Scientific, Singapore, 1985).

 \bibitem{fs} See, e.g., http://en.wikipedia.org/wiki/Fubini-Study\_metric.

 \bibitem{PO} See, e.g., http://en.wikipedia.org/wiki/Poincar\'e\_ half-plane\_model.

\bibitem{kl3} J.R.~Klauder and E.W.~Aslaksen, ``Elementary Model for Quantum
        Gravity", Phys. Rev. D {\bf 2}, 272-276 (1970).

\bibitem{fz} M.~Fanuel and S.~Zonetti, ``Affine Quantization and the Initial Cosmological Singularity'',
     EPL {\bf 101},  10001 (2013); arXiv:1203.4936.

 \bibitem{rs1}J.R.~Klauder, ``Rotationally-Symmetric Model Field
Theories", J. Math. Phys. {\bf 6}, 1666-1679 (1965).

\bibitem{rs2} J.R.~Klauder, {\it Beyond Conventional Quantization}, (Cambridge Uniiversity Press, Cambridge, 2000 \& 2005).

\bibitem{rs3} H.D.I.~Abarbanel, J.R.~Klauder, and J.G.~Taylor, ``Green's Functions for Rotationally Symmetric Models", Phys. Rev. {\bf 152}, 198-206 (1966).

\bibitem{kl6}  J.R.~Klauder, ``Enhanced Quantization: A Primer'',  J. Phys. A: Math. Theor. {\bf 45}, 285304 (8pp) (2012); arXiv:1204.2870.

\bibitem{bg} J.~Ben~Geloun and J.R.~Klauder, ``Enhanced Quantization on the Circle'', Phys. Scr. {\bf 87},  035006 (5pp) (2013); arXiv:1206.1180.   J.~Ben~Geloun, ``Enhanced Quantization: The Particle on the Circle''; arXiv:1210.5492.

\bibitem{kl7} J.R.~Klauder, ``The Utility of Affine Variables and Affine Coherent States'', J. Phys. A: Math.
Theor. {\bf 45}, 24001 (17pp) (2012); arXiv:1108.3380.

\bibitem{kll} J.R.~Klauder  ``Enhanced Quantum Procedures that Resolve Difficult Problems'', arXiv:1206.4017.

\bibitem{k33} J.R.~Klauder, ``Divergences in Scalar Quantum Field Theory: The Cause and the Cure'',
     Mod. Phys. Lett. A {\bf 27}, 1250117 (9pp) (2012); arXiv:1112.0803.

\bibitem{MM} J.R.~Klauder, ``Matrix Models and their Large-N Behavior'', Int. J. Mod. Phys. A {\bf 29}, 1450026 (14 pages) (2014); arXiv:1312.0814.

\bibitem{AK} T. C. Adorno and J. R. Klauder, ``Examples of Enhanced Quantization: Bosons, Fermions, and Anyons'', arXiv:1403.1786

\bibitem{KL} J.R.~Klauder, ``An Affinity for Affine Quantum Gravity'',
Proceedings of the Steklov Institute of Mathematics {\bf 272}, 169 - 176 (2011); arXiv:1003.2617.

\bibitem{kl8} J.R.~Klauder, ``Recent Results Regarding Affine Quantum Gravity'', J. Math. Phys. {\bf 53}, 082501 (2012) (19pp); arXiv:1203.0691.


       \end{thebibliography}
\end{document}